\documentclass[11pt,reqno,a4paper]{amsart}
\usepackage{graphicx,epsfig}
\usepackage{amsaddr}
\usepackage{enumerate}
\usepackage[final,color,notref,notcite]{showkeys}
\usepackage{caption}
\usepackage{todonotes}
\usepackage{soul}
\usepackage{hyperref}
\usepackage[numbers,sort&compress]{natbib}
\usepackage[charter]{mathdesign}
\usepackage[scr=boondoxo, scrscaled=.98]{mathalfa}

\newtheorem*{thm}{Theorem}
\DeclareMathAlphabet{\pazocal}{OMS}{zplm}{m}{n}   
\newcommand{\Ccal}{\pazocal{C}}
\newcommand{\Ncal}{\pazocal{N}}
\newcommand{\Ecal}{\pazocal{E}}
\newcommand{\Dcal}{\pazocal{D}}
\newcommand{\Mcal}{\pazocal{M}}

\newcommand{\Bcal}{\pazocal{B}}
\newcommand{\Gcal}{\pazocal{G}}

\newcommand{\ket}[1]{\mathop{|#1\rangle}\nolimits}

\newcommand{\ave}[1]{\langle#1\rangle}

\newcommand*{\at}{@}

\definecolor{labelkey}{rgb}{0,.56,.7}

\def\bbZ{\mathbb{Z}}
\def\bbR{\mathbb{R}}
\newcommand{\nn}{\nonumber}
\def\wh{\widehat}
\def\dg{\dagger}
\def\df{\overset{\mathrm{df}}{=}}
\newcommand{\Tr}[1]{\mathop{{\mathrm{Tr}}_{#1}}}
\newcommand{\id}{\mathop{{\mathrm{id}}}\nolimits}
\newcommand{\rank}{\mathop{{\mathrm{rank}}}\nolimits}
\DeclareMathOperator\acos{arccos}

\def\a{\alpha}
\def\b{\beta}
\def\g{\gamma}
\def\d{\delta}

\def\ve{\varepsilon}
\def\vr{\varrho}
\def\vp{\varphi}
\def\ka{\kappa}
\def\om{\omega}
\def\Om{\Omega}
\def\s{\sigma}

\def\vt{\vartheta}

\def\om{\omega}
\def\Om{\Omega}

\sodef\so{}{.065em}{.4em plus1em}{2em plus.1em minus.1em}

\begin{document}

\title[Black holes as bosonic Gaussian~channels]{\Large\so{Black holes as bosonic Gaussian~channels}}

\begin{abstract}
    We identify the quantum channels corresponding to the interaction of a Gaussian quantum state with an already formed  Schwarzschild black hole. Using recent advances in the classification of one-mode bosonic Gaussian channels we find that (with one exception) the black hole Gaussian channels lie in the {\em non-entanglement breaking} subset of the lossy channels $\Ccal(\mathrm{loss})$, amplifying channels $\Ccal(\mathrm{amp})$ and classical-noise channels $\Bcal_2$. We show that the channel parameters depend on the black hole mass and the properties of the potential barrier surrounding it. This classification enables us to calculate the classical and quantum capacity of the black hole and to estimate the quantum capacity where no tractable quantum capacity expression exists today. We discuss these findings in the light of the black hole quantum information loss problem.
\end{abstract}


\author{Kamil Br\'adler}
\email{kbradler\at ap.smu.ca}
\address{
    Department of Astronomy and Physics,
    Saint Mary's University,
    Halifax, Nova Scotia, B3H 3C3, Canada
    }

\author{Christoph Adami}
\address{Department of Physics and Astronomy, Michigan State University, East Lansing, MI 48824}

\maketitle
\thispagestyle{empty}

\section{Introduction}
\begin{figure}[htbp]
   \centering
   \includegraphics{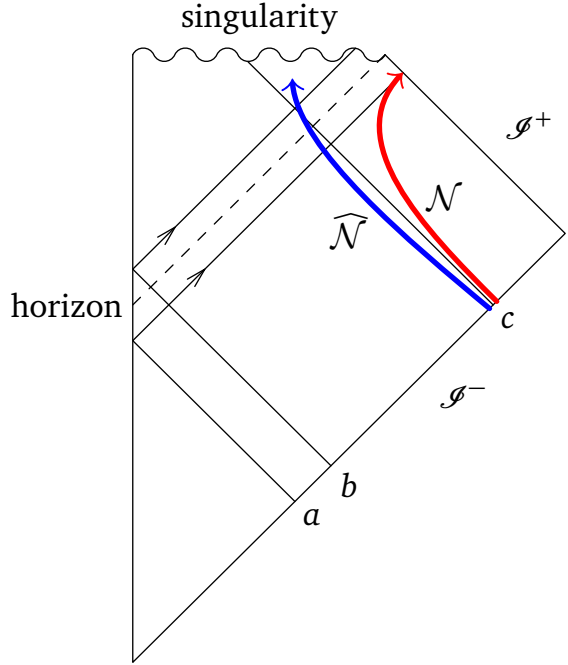}
   \caption{The modes $a$, $b$, and $c$ are concentrated in a region of past null infinity indicated by the letter. Modes $a$ and $b$ are the standard early-time modes that travel just outside and just inside the horizon, while mode $c$ is a late-time mode that is severely blue-shifted with respect to the $a$ and $b$ modes. The operator $A$ annihilates modes outside the horizon at future infinity. The channels $\Ncal$ (red) and its complement $\wh\Ncal$ (blue) are also indicated. }
   \label{fig:sorkin}
\end{figure}

Because the standard semi-classical approximation to quantum gravity is a free-field theory in curved space-time~\cite{hawking1975particle}, there is currently no unambiguous way to introduce interactions between radiation and black holes~\cite{Witten1992} (but see~\cite[Chap.9]{BirrellDavies1982} and ~\cite{HollandsWald2010} for methods to introduce interactions in an axiomatic way). Yet, the problem of information loss in black holes~\cite{Hawking1976b,Preskill1992,Giddings1996,Mathur2009} (as well as the related firewall paradox~\cite{braunstein2009black,braunstein2013better,almheiri2013black}) explicitly considers the fate of information-bearing particles interacting with a black hole.  A way out of this conundrum has recently been proposed, by explicitly studying the interaction of a scalar massless field with an already formed Schwarzschild black hole in terms of quantum channel theory~\cite{bradler2013capacity}. Such an interaction can be written down using Sorkin's effective model~\cite{sorkin1987simplified}, which makes explicit the interactions described implicitly by the grey-body factor of Hawking~\cite{hawking1975particle}, and in particular allows for a calculation of the spectrum of radiation emitted by a black hole in response to late-time incoming radiation (such a calculation was first presented for early-time incoming states by Bekenstein and Meisels~\cite{BekensteinMeisels1977} as well as Panangaden and Wald~\cite{PanangadenWald1977}  (see also~\cite{AudretschMueller1992}). In Sorkin's construction, the outgoing field operator $A$ is related to the incoming modes by a Bogoliubov transformation
 \begin{equation}\label{eq:Bogoliubov}
  A=\a a+\b b^\dg+\g c\;.
\end{equation}
Here,  $a,b$ and $c$ are the annihilation operators defining the early- and late-time particle content in the incoming sector respectively, and $\a,\b,\g$ are coefficients to be determined later. The early-time modes $a$ and $b$ are associated with quantum fields that were emitted during the formation of the black hole and travel just inside ($b$) and just outside ($a$) the event horizon, are standard within the literature of curved space quantum field theory.  As Hawking showed~\cite{hawking1975particle}, when propagated toward future null infinity these horizon modes are exponentially red shifted relative to the frequencies that a stationary observer at late time might expect. If that observer would send her own $c$ modes into the black hole, the relative blue-shift of these modes with respect to the black hole horizon modes implies that the support of the quantum fields associated with $c$ modes is disjoint from that of the $a$ and $b$ modes. As a consequence, the outgoing field operator $A$ should resolve into a superposition not just of the ingoing horizon modes $a$ and $b$, but also the ingoing late-time blue-shifted ``signal'' modes $c$~\cite{sorkin1987simplified}, see Fig.~\ref{fig:sorkin} (note that our notation differs from~\cite{sorkin1987simplified}).

Using the expanded Bogoliubov transformation~(\ref{eq:Bogoliubov}) Sorkin showed that the resulting expression for the radiation experienced by a stationary observer suspended far away from the black hole horizon precisely reproduces the standard Hawking radiation effect \emph{ including} the effect of a black hole potential (grey-body factor) whose parameters are implicit in the coefficients in Eq.~(\ref{eq:Bogoliubov}). Adami and Ver Steeg~\cite{AdamiVersteeg2013} then recently showed that the Bogoliubov transformation~(\ref{eq:Bogoliubov}) is in fact completely analogous to a corresponding relation in quantum optics, with an interaction term between the late-time modes $c$ and early-time horizon modes $a$ implemented by a beam-splitter Hamiltonian (see also~\cite{AdamiVersteeg2015}). Thus, Sorkin's construction enables a direct analysis of the interaction of scalar massless particles with a black hole horizon, and makes it possible to investigate the capacity to transmit classical or quantum information via a black hole.

The operator relation~Eq.~(\ref{eq:Bogoliubov}) completely characterizes the evolution of any input (bosonic) state, and therefore is sufficient to study the fate of quantum information. However, the bosonic sector of the (infinite-dimensional) Fock space is unwieldy, hence it is often advisable to further limit the input Hilbert space to a physically motivated subset. One option investigated in~\cite{AdamiVersteeg2013,bradler2013capacity} is to confine the input Fock space to be a sector spanned by a vacuum $\ket{0}$ and a single photon state $\ket{1}$. Then, an arbitrary two-level state (qubit) can be constructed using the so-called dual-rail encoding~\cite{nielsen2010quantum}.

Here we examine instead a different input subset that is a favorite choice in quantum optics due to its experimental relevance -- Gaussian states. Gaussian states are completely described by the first two moments of the canonical quadrature variables. The most prominent examples of Gaussian states are coherent states, (multi- or single mode) squeezed states, and thermal states. Focusing on Gaussian states narrows the number of possible input states substantially without unduly simplifying the system. Owing to the form of~Eq.~(\ref{eq:Bogoliubov}), classifying the black hole quantum channel is reduced to the study of one-mode Gaussian (OMG) channels~\cite{holevo2012quantum} (loosely defined as bosonic completely positive maps transforming Gaussian states into Gaussian states). The set of all possible OMG channels has recently been exhaustively classified~\cite{holevo2007one,holevo2012quantum,weedbrook2012gaussian}. We now know that there exist eight equivalence classes of OMG channels that are embodiments of various passive and active optical elements. Among those channels, we find for example the family of \emph{ lossy channels} (implemented by imbalanced beam splitters)  as well as the \emph{ amplification} and \emph{ conjugated} channel families, related to the  parametric amplification process. For details on those channels, we refer the reader to the excellent expositions~\cite{holevo2012quantum,weedbrook2012gaussian}.

OMG channels are important in another respect: for lossy and amplifying OMG channels with added classical noise  (sometimes called phase-insensitive Gaussian channels) it is possible to calculate their \emph{ classical  capacity}~\cite{giovannetti2013ultimate}, that is, the capacity to transmit classical information via a quantum channel. For other situations where the classical capacity is calculable see~\cite{schafer2013equivalence}. Here we  calculate the \emph{ quantum} capacities of these OMG channels. Viewing the black hole in terms of the OMG channel construction will allow us to study how much information -- encoded in a Gaussian state and sent into the black hole horizon -- can be recovered by an outside stationary observer. Understanding what happens to information (both classical and quantum) incident on a black hole is known as the black hole information problem (even though what is and is not a problem is often hotly contested~\cite{mathur2011information}). A rigorous analysis of this problem must begin with the identification of the physical system (incident states and black hole) in terms of a quantum channel, and calculating its channel capacities as argued in~\cite{AdamiVersteeg2013,bradler2013capacity}. As information theory has not yet become a standard tool in the relativist's arsenal, we present in the next section the minimal background needed to appreciate the power of channel capacity theorems, in particular their quantum realizations.

\section{The quantum capacity of a noisy quantum channel: Construction and Interpretation}

The success of quantum Shannon theory developed in the past fifteen years draws from its classical counterpart, created virtually singlehandedly by Shannon~\cite{shannon2001mathematical,cover2012elements} Shannon's theory is often called \emph{the} mathematical theory of classical communication and as its name suggests, one of its main goals is to rigorously calculate the rate at which two or more parties can reliably communicate even under presence of noise. The principal idea of how to achieve it is to add some redundancy to the message and craft it in such a way that the noise ``eats up'' the added redundancy and leaves the message intact.  The message transformation is called \emph{encoding} and its outcome is a \emph{code}, so the purpose of the encoding is to make the message error-correctable. The encoding must ensure that the received message can be arbitrarily perfectly recovered (decoded) and that the amount of redundancy must be as small as possible. In other words, it must be possible to substitute the whole noisy transmission line with a noiseless one at the lowest cost. The quantity that characterizes the efficiency of this algorithm is called the \emph{noisy channel capacity} and Shannon's major achievement in this direction was the derivation of a simple capacity formula for a fairly general model of classical communication. The capacity is essentially the maximal ratio of the message length to the code length given the amount of noise. Two extreme examples serve to illustrate the notion: if the channel capacity is maximal, there is no need to create a code as the channel is already noiseless. If the capacity is zero, there is no way of correcting the errors and reliably conveying \emph{ any} classical information -- the channel is useless for communication purposes. For any value in between, a larger capacity implies a faster way to transmit the information perfectly (even though the precise manner of encoding to reach the capacity might not be known). It is important to note that if the capacity of a channel is non-zero, then it is guaranteed that it is possible to transmit information with arbitrary accuracy, but it could take very long to do so.

Many aspects of the classical capacity concept carry over to the quantum case but in general the theory is far richer, with many questions still open. Quantum mechanics is obviously a more general theory than classical physics, allowing for a large variety of information transmission modes. It is possible to transmit classical or quantum bits through a quantum channel, while having classical or quantum resources at disposal.  There is one crucial difference between classical and quantum information theory that is worth highlighting: Unlike classical Shannon theory (that could be classified as part of engineering), the quantum conveys something truly fundamental about a physical system. It tells us how.

One aspects that carries over to quantum Shannon theory is its asymptotic character.  (The quantum theory dealing with the single use of a channel -- the ``one shot capacity'' -- is currently under vigorous development~\cite{tomamichel2015quantum}). In the asymptotic regime it is assumed that the channel can be used $n$ times and the capacity results are derived under the assumptions of $n\to\infty$. This is normally a physically reasonable assumption and also the first step towards usually more involved one-shot settings, where the channel can be used only once or twice, for instance. The commutative asymptotic theory relying on the notions of \emph{typical sets} was already used by Shannon~\cite{cover2012elements} and was relatively recently generalized to the quantum setting by taking into account the inherent noncommutativity of the quantum world~\cite{jozsa1994new,holevo2012quantum}.

\begin{figure}[t]
  \centering
  \includegraphics{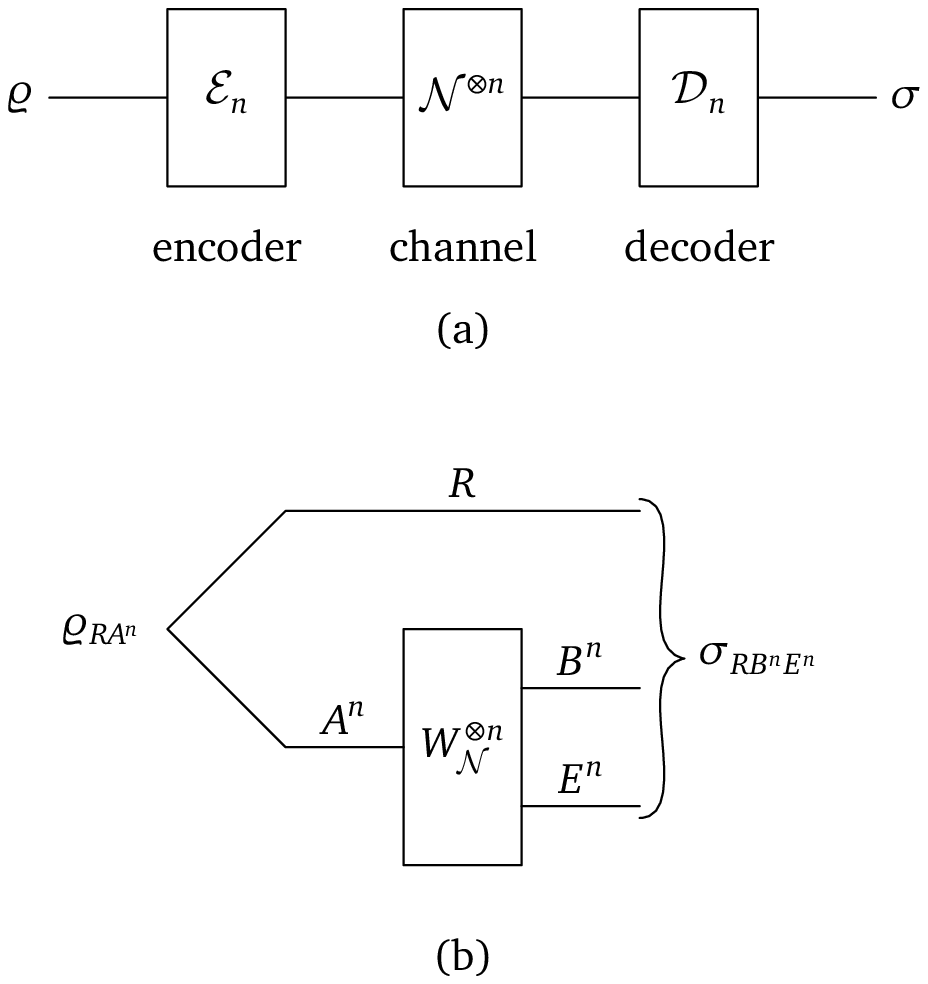}
     \caption{Quantum communication channels. (a) A  setup where the task is to reliably transmit an arbitrary state $\vr$ from a sender to a receiver through $n$ copies of noisy quantum channel $\Ncal$. If the channel is not too noisy the encoder adds enough redundancy so that the errors can be corrected by the recipient and the quantum state $\vr$ can be reconstructed. The output state $\sigma$ is, in the asymptotic limit $n\to\infty$, indistinguishable from $\varrho$. (b) An equivalent quantum channel using the purified picture, that simplifies the formulation of the direct coding theorems for the quantum capacity of a noisy channel.}
   \label{fig:eccn}
\end{figure}
In quantum Shannon theory (see for example~\cite{holevo2012quantum,Wilde2013}) the goal is exactly the same as in the classical framework: the intention is to simulate a noiseless quantum channel by attaching an encoder $\Ecal$ to the noisy communication channel $\Ncal$ followed by a decoder $\Dcal$ on the receiver's side: $\id\simeq\Dcal_n\circ\Ncal^{\otimes n}\circ\Ecal_n$ (see Fig.~\ref{fig:eccn}(a)). Again, we ask for the composite channel to be arbitrarily close (using an operationally motivated distance measure introduced later) to the identity as $n$ grows. Given the quantum character of the physical system, we may decide whether to send classical or quantum information over the channel. The classical capacity of a noisy quantum channel quantifies how much \emph{classical} information (encoded in a quantum state) can be perfectly recovered from the output of the channel. Similarly, the quantum channel capacity quantifies the amount of \emph{quantum} information that can be transmitted with a vanishing error. Both channel capacities are fundamental quantities -- they represent the maximal rates at which the respective information can be transmitted through a quantum system. Given the nature of the physical system studied in this paper (a Schwarzschild black hole), we will be interested in transmitting quantum information through a quantum channel. Semiclassical  black holes are often accused of ``losing'' information, but often the term ``information'' is not well defined in the context. Indeed, when the black hole information problem was first formulated, the concepts of classical and quantum capacity of quantum channels did not even exist. But they do now, and so it behooves us to study the fate of quantum information that is either already present in the black hole~\cite{BradlerAdami2015} or is sent toward the horizon after the blackhole is already formed, using the appropriate framework.

\subsection{Direct coding theorem for entanglement transmission over a noisy quantum channel}
\label{subsec:directcode}

We review here certain crucial aspects of the derivation of the quantum capacity in the asymptotic case, but in a modern reformulation. Quantum information transmission can be conceived in terms of two at first sight different communication setups, that however turn out to be equivalent. The most straightforward formulation is depicted in Fig.~\ref{fig:eccn}(a) where a sender prepares an arbitrary quantum state $\vr$ and sends it through a noisy quantum channel $\Ncal$, in the hope that the message can be recovered by the receiver via error correction using the decoder. It turns out that this formulation is hard to analyze, but a crucial reformulation of this process in terms of the ``purified'' system (shown in Fig.~\ref{fig:eccn}(b)) allows much progress. In the process depicted in Fig.~\ref{fig:eccn}(b), part of an entangled state is sent through the channel, but such a scenario is in fact equivalent given the insight that such an entangled state will allow the transmission of arbitrary quantum states via quantum teleportation~\cite{Bennettetal1993}.

A state $\vr_A$ is ``purified'' by $\vr_{RA}$ if $\vr_A=\Tr{R}\vr_{RA}$ and $\vr_{RA}$ is pure. A quantum channel can be purified as well, on account of to the Stinespring theorem~\cite{paulsen2002completely} that states that for every completely positive map $\Ncal$ there exists a partial isometry $W_\Ncal$ such that $\Ncal(\vr_A)=\Tr{E}[W_\Ncal\vr_A W_\Ncal^\dg]$. The isometry thus ``lifts''  the quantum channel to a larger Hilbert space, also known as the Stinespring dilation. Recall that an isometry $V$ is a linear map between two normed vector spaces $A$ and $B$ such that $|\zeta|_A=|V\zeta|_B$ for all $\zeta\in A$, where $|\cdot|$ denotes a norm (this is equivalent to $V^\dg V=\id$). A partial isometry $W$ is a linear map whose restriction to the complementary subspace of its kernel is an isometry, and it is a projector: $(W^\dg W)^2=W^\dg W$. Hence instead of Fig.~\ref{fig:eccn}(a) we will  treat the equivalent process depicted in Fig.~\ref{fig:eccn}(b). Tracing over the output $B$ of the isometric extension $W_\Ncal$ induces a supremely important notion of a \emph{complementary channel} to $\Ncal$: $\wh\Ncal(\vr_A)\df\Tr{B}[W_\Ncal\vr_A W_\Ncal^\dg]$ ~\cite{holevo2012quantum}. Note that isometries and purifications are not unique\footnote{Also note that for infinite-dimensional Hilbert spaces the operator theory becomes substantially more complicated. By restricting to ourselves to Gaussian states and operators, many of those problems are avoided~\cite{holevo2001evaluating}.} 

\begin{figure}[t]
  \centering
  \includegraphics{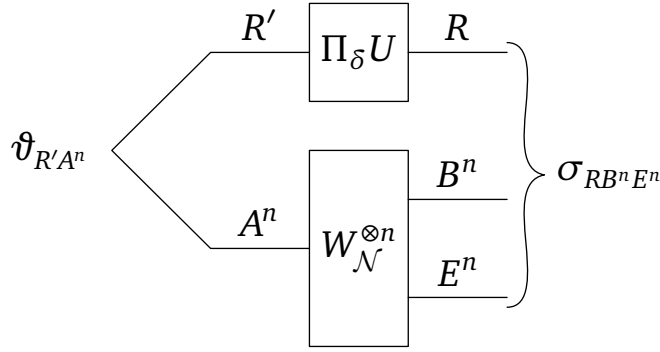}
     \caption{Sketch of the setup for decoupling via random encoding. The goal is to transmit part of an entangled state $\ve_{R'A^n}$ through the $n$ copies of the channel $\Ncal$ by decoupling the $RB^n$ subsystem from the complementary channel output $E^n$. The random code is a random matrix $\Pi_\d U$, where $U$ is a randomly and uniformly chosen unitary operator and $\Pi_\d$ is a projector onto a $\d$-typical Hilbert subspace.  This explicit but impractical procedure achieves the decoupling whose precise formulation is shown in Eq.~(\ref{eq:DeltaDecoupling}) and below.}
   \label{fig:decouple}
\end{figure}

Before defining the quantum channel capacity formally, we state the \emph{ direct coding theorem} of quantum information theory (the direct theorem proves the existence of a coding scheme that achieves the capacity) due to~\cite{kretschmann2004tema,devetak2005private}, but in a modern reformulation of the problem. Hayden et al.'s approach~\cite{hayden2008decoupling} based on a \emph{ decoupling theorem} (see below)  is conceptually clearer and arguably superior to the original formulation in that it allows for a generalization to the one-shot setting~\cite{dupuis2014one}, going well beyond mere reliable quantum communication over a noisy quantum channel.

The decoupling theorem states that for a randomly and uniformly chosen unitary operator $U$, the following inequality holds:
\begin{equation}\label{eq:DeltaDecoupling}
\int_U\big\|\s^{(\d)}_{RE^n}-I_{R}\otimes\,\s^{(\d)}_{E^n}\big\|_1\mathrm{d}U\leq2^{\frac{n}{2}[K-H(B)_\s+H(E)_\s+2\d]}\leq\ve_n.
\end{equation}
The integrand is a unitary expectation value of a distance between two states, induced by the trace norm $\|A\|_1\df\Tr{}[(A^\dg A)^{1/2}]$ (the Schatten 1-norm). The theorem states that a randomly chosen unitary $U$ causes the state $\s^{(\d)}_{RE^n}$ to become close to a product of a (normalized) identity $I_R$ and $\s^{(\d)}_{E^n}$ as long as the right hand side of (\ref{eq:DeltaDecoupling}) is made sufficiently small by the choice of the \emph{ rate} $K$, where $2^{nK}=\dim{R}$. The symbol $\d$ stands for states from a $\d$-typical subspace of a tensor product of many copies of a Hilbert space. A $\d$-typical Hilbert space is a generalization of a $\d$-typical set from classical Shannon theory (see \cite{cover2012elements,hayden2008decoupling} for the proper classical and quantum definition).

We note the appearance of the von Neumann entropy $H(B)_\s\df-\Tr{}[\s_B\log{\s_B}]$ (the entropy of the final state density matrix, see Fig.~\ref{fig:decouple}) in the exponent of Eq.~(\ref{eq:DeltaDecoupling}).  Together with $H(E)_\s$ (the entropy of the environment), they make up the  \emph{coherent information}~\cite{barnum1998information}
\begin{equation}\label{eq:DEFcohinfo}
      \tilde I_{\mathrm{coh}}(\Ncal,\vt_A)\df H(B)_\s-H(E)_\s.
\end{equation}
The decoupling theorem states that whenever $0\leq K< I_{\mathrm{coh}}(\Mcal,\vt_A)$, the right hand side of (\ref{eq:DeltaDecoupling}) goes to zero and decoupling (of the system $A$ from its entangled partner $R$) follows.

It is possible to show that decoupling for $\d$-typical states in Eq.~(\ref{eq:DeltaDecoupling}) implies decoupling for \emph{ all} states~\cite{hayden2008decoupling} and, crucially, decoupling holds for almost all unitaries:
\begin{equation}\label{eq:decoupling}
  \big\|\s_{RE^n}-I_{R}\otimes\,\s_{E^n}\big\|_1\leq\tilde\ve_n.
\end{equation}
Even more remarkably, due to decoupling the owner of the $B$ subsystem (the receiver) is able to decode and prepare a maximally entangled state
\begin{equation}\label{eq:maxent}
    \Phi_{RD}=2^{-nK/2}\sum_{i=1}^{2^{nK}}\ket{i}_R\ket{i}_{D}
\end{equation}
of dimension given by the rate $K$, where the $D$ subscript is the output of the receiver's decoder $\Dcal_n$. In  other words, the sender and receiver are able to perfectly transmit quantum information through the channel $\Ncal$ at the rate $K$ of the entanglement transmission protocol. 

To formally define the quantum capacity,  we need to introduce some additional terminology. Using the definition of the \emph{fidelity} between a mixed state $\vr$ and a pure state $\vp$ given as $F(\vp,\vr)\df\langle\vp|\vr|\vp\rangle$, we further observe:
\begin{enumerate}[(I)]
  \item The pair $(\vt_{R'A^n},\Dcal_n)$ is called a $(K,n,\tilde\ve_n)$ \textbf{entanglement generation code} for the channel $\Ncal$ if
  \[F(\Phi_{RD},\Dcal_n\circ\Mcal^{\otimes n}\circ\vt_{R'A^n})\geq1-\tilde\ve_n.\]
  \item A rate $K$ is \textbf{achievable} if there is a sequence of $(K,n,\tilde\ve_n)$ such that $\tilde\ve_n\to0$.
  \item The \textbf{quantum channel capacity} $Q(\Ncal)$ is the maximum over all achievable rates.
\end{enumerate}
We can now state the direct coding theorem in terms of decoupling theory as follows~\cite{hayden2008decoupling}:
\begin{thm}[Direct coding theorem]
For a channel $\Ncal$ and a code $\vt_A$ every rate $K$ is achievable as long as $0\leq K<\tilde I_{\mathrm{coh}}(\Ncal,\vt_A)$.
\end{thm}
The single-copy quantum capacity (the optimized coherent information) is then
\begin{equation}\label{eq:singlecopyQcap}
    I_{\mathrm{coh}}(\Ncal)\df\sup_{\vt_A}{\tilde I_{\mathrm{coh}}(\Ncal,\vt_A)}.\;.
\end{equation}
By ``bundling''  $k$ copies of $\Ncal$ at the same time (setting $mk=n$, so that $m\to\infty$) and using some elementary properties of the coherent information we can show that
\begin{equation}\label{eq:kcopies}
0\leq K< \frac{1}{k}\tilde I_{\mathrm{coh}}(\Ncal^{\otimes k},\vt_{A^k}).
\end{equation}
By increasing $k$ we finally arrive at a lower bound for the \emph{quantum capacity formula}:
\begin{equation}\label{eq:QuantCaplowerbound}
Q(\Ncal)\geq\lim_{k\to\infty}{\frac{1}{k}\sup_{\vt_{A^k}}\tilde I_{\mathrm{coh}}(\Ncal^{\otimes k},\vt_{A^k})}\;.
\end{equation}
This is almost the final result if we invoke the \emph{converse theorem} (the reverse of the {\rm direct theorem}) that happens to prove Eq.~(\ref{eq:QuantCaplowerbound}) with the opposite inequality\footnote{The importance of the converse theorem lies in showing the iff condition between the achievability and the rate smaller than the capacity.} leading to the  central result for the quantum capacity:
\begin{equation}\label{eq:QuantCap}
Q(\Ncal)=\lim_{k\to\infty}{\frac{1}{k}\sup_{\vt_{A^k}}\tilde I_{\mathrm{coh}}(\Ncal^{\otimes k},\vt_{A^k})}\;.
\end{equation}
The latter is called a \emph{multi-letter} capacity formula, and it disappointing in the sense that this regularized expression is incalculable because the infinite limit makes the optimization intractable. To some extent, this is the crucial difference between classical and quantum information theory. Quantum mechanics allows the use entangled states $\vt_{A^k}$ as codes, and it is not known over what set of states to optimize the capacity, or even how to parameterize it. Even if we fixed $k$ to be a reasonably small integer, we would struggle to find the maximum in the above expression. As opposed to classical Shannon theory where single-letter (no regularization) results are abundant (an example being Shannon's celebrated results mentioned earlier) quantum Shannon theory generically produces multi-letter capacity formulas that are not calculable.

A consequence of this intractability is that the quantum capacity is currently known only for three special cases of channels called \emph{degradable}~\cite{devetakshor2005capacity}, \emph{conjugate degradable}~\cite{bradler2010conjugate} and \emph{antidegradable} (the quantum capacity vanishes for antidegradable channels) and in general only a lower bound can be given using Eq.~(\ref{eq:singlecopyQcap}). We recall that a channel $\Ncal$ is degradable if there exists another channel $\Gcal$ (called a degrading channel) such that $\Gcal\circ\Ncal=\wh\Ncal$, where $\wh\Ncal$ is a complementary channel to $\Ncal$. In this case $\wh\Ncal$ is called antidegradable. The complementary channel captures the evolution of the channel's environment, and so degradable channels can simulate that environment by composition with the degrading map. Hence, for degradable channels there exists a single-letter quantum capacity formula
\begin{equation}\label{eq:singleletter}
  Q(\Ncal)=I_{\mathrm{coh}}(\Ncal).
\end{equation}
Degradable and antidegradable channels will play an important role in the Gaussian scenario~\cite{holevo2001evaluating,wolf2007quantum} studied here (see Sec.~\ref{sec:complementarychannels}), allowing us
 to estimate the quantum channel capacity for a substantial fraction of the OMG channels that occur in black holes, and calculate it exactly in certain special cases.

In order to understand the flow of quantum information we also extensively study the complementary quantum channel, corresponding to the transmission of quantum information \emph{behind} the horizon (see Fig.~\ref{fig:sorkin}), to be decoded by a potential observer \emph{inside} the black hole. We find that the process of absorption of quantum states by the black hole horizon is, in the language of OMG channels, described by the family of \emph{conjugated channels}. We discuss the implications for the transmission of quantum information in the concluding section.

\section{Black hole as an OMG channel}
\label{sec:OMG}
While not stated explicitly by Sorkin, we show below that the coefficients in Eq.~(\ref{eq:Bogoliubov}) describing the field dynamics in the Heisenberg picture belongs to the symplectic group over the reals $Sp(6,\bbR)$, and the coefficients in Eq.~(\ref{eq:Bogoliubov}) satisfy $|\a|\leq1$ and $\b,\g\in\bbR$. Moreover, the coefficients must be such that the outgoing field operator is correctly normalized:~$[A,A^\dg]=1$. In order to derive these conditions we set\footnote{Note that this parametrization is different from the ansatz made in~\cite{AdamiVersteeg2013} that, however, leads to the same solution.}:
\begin{equation}\label{eq:ansatz}
  X=r(\ka a^\dg b^\dg-\ka ab)+s(a^\dg c-ac^\dg),
\end{equation}
and identify $A=e^Xae^{-X}$ with the right-hand side of Eq.~(\ref{eq:Bogoliubov}). In~(\ref{eq:ansatz}) we recognize the Hamiltonians for a two-mode squeezing transformation as well as a beam-splitter, where $r,s,\ka>0$. We restricted the Hamiltonian parameters to be real in hindsight, as complex phases will not play any role. The Baker-Campbell-Hausdorff theorem  leads to an identification of the  coefficients in Eq.~(\ref{eq:Bogoliubov}) with the Hamiltonian parameters as we will exemplify on the $A$ operator relation. We first rewrite Eq.~(\ref{eq:ansatz}) as follows
\begin{align}\label{eq:ArelationDerivation}
  X  =    r a^\dg\Big(\ka b^\dg+{s\over r}c\Big)-ra\Big(\ka b+{s\over r}c^\dg\Big)
    \equiv ra^\dg d-rad^\dg.
\end{align}
We thus find
\begin{equation}\label{eq:explicitcalc}
  A=e^Xae^{-X}=\cos{r}\,a-\sin{r}\,d=\cos{r}\,a-\sin{r}\Big(\ka b^\dg+{s\over r}c\Big)
\end{equation}
and by asking $[A,A^\dg]=1$ we find that $\ka^2=s^2/r^2-1$ must be satisfied. Note that this is the same condition given by $[d,d^\dg]=1$. Hence we obtain
\begin{subequations}\label{eq:abg}
  \begin{align}
    \a & = \cos{r},\label{eq:abgI}\\
    \b & = \pm\left({s^2\over r^2}-1\right)^{1/2}\sin{r},\\
    \g & = -{s\over r}\sin{r}.
  \end{align}
\end{subequations}
Later (see below Eqs.~(\ref{eq:tauANDy})) we will find that the requirement on complete positivity of a derived OMG channel further imposes $s\geq r$.

A similar procedure for the out field-operators $B=e^Xbe^{-X}$ and $C=e^Xce^{-X}$ gives rise to the following transformation of a column list of field operators~$L:\{a,b^\dg,c\}\mapsto\{A,B^\dg,C\}$:
\begin{equation}\label{eq:SU11}
L=\begin{bmatrix}
     \cos{r} & -\left({s^2\over r^2}-1\right)^{1/2} \sin{r} & -\frac{s  \sin{r}}{r} \\
     -\left({s^2\over r^2}-1\right)^{1/2} \sin{r} & \frac{s ^2+\left(r^2-s ^2\right) \cos{r}}{r^2} & -\frac{s  \left({s^2\over r^2}-1\right)^{1/2} (\cos{r}-1)}{r} \\
     \frac{s  \sin{r}}{r} & \frac{s  \left({s^2\over r^2}-1\right)^{1/2} (\cos{r}-1)}{r} & \frac{(\cos{r}-1) s ^2}{r^2}+1 \\
\end{bmatrix}.
\end{equation}
By introducing a symplectic form
\begin{equation}\label{eq:symplecForm}
  \om=\begin{bmatrix}
        0 & 1 \\
        -1 & 0 \\
      \end{bmatrix}
\end{equation}
and setting $\Om\df\om\oplus\om\oplus\om$ we find that $(L\oplus L)\Om(L\oplus L)^T=\Om$ for $L\oplus L$ acting on a column list of field operators assembled as $\{a,a^\dg,b,b^\dg,c,c^\dg\}$. Hence $L\in Sp(6,\bbR)$. We now would like to write $L$ in the basis of quadrature operators $p$ and $q$~\cite{dutta1994squeezed} given by $\s:\{q,p\}\mapsto\{a,a^\dg\}$ where
\begin{equation}
\s=
\begin{bmatrix}
  1 & i \\
  1 & -i \\
\end{bmatrix}
\end{equation}
and we set $\hbar=2$. This can be achieved by writing the global evolution transformation of quadrature operators as
\begin{equation}\label{eq:Smatrix}
  S=\Sigma^{-1}(L\oplus L)\Sigma,
\end{equation}
where $\Sigma\df\s\oplus\s\oplus\s$ and one can verify that $S\Om S^T=\Om$ is satisfied (for more information see for example~\cite{ferraro2005gaussian,adesso2007entanglement}). The $S$ matrix introduced above will allow us to find the black hole response to an arbitrary incoming Gaussian state (for comprehensive definitions of Gaussian states and transformations see~\cite{holevo2012quantum}).

On physical grounds, we are interested in the following scenario. The early-time (input) black hole horizon modes ($a$ and $b$) are in a vacuum state and the late-time incoming mode $c$ will be a one-mode Gaussian state that is completely described by a  covariance matrix $\mathbf{V}_{\mathrm{in}}$ capturing the second quadrature moments that completely determines the channel. Hence we can  set the first moments describing the state displacement in phase space to zero. Because $S$ is a symplectic transformation, the output modes are Gaussian. We are not, however, interested in the outgoing states per se. Our task is to deduce from the output covariance matrix $\mathbf{V}_{\mathrm{out}}$ which of the OMG channels described in the introduction is responsible for the state transformation.

To study this we use the recent complete classification of OMG channels~\cite{holevo2007one}. It was found that there are eight equivalence classes of OMG channels~\cite{holevo2007one,holevo2012quantum,weedbrook2012gaussian} (modulo two Gaussian unitaries -- one preceding and the other following the channel) distinguished by the values of three parameters that fully characterize the OMG completely positive maps. The Stinespring dilations relevant for the OMG channels studied in this paper are two-mode symplectic transformation with a complementary (reference) input given
by a thermal state of mean photon number $\ave{n}$. This number is the first of the three parameters  characterizing the OMG channels. The two remaining parameters come from a generic evolution of one-mode covariance matrices
\begin{equation}\label{eq:CMevolution}
  \mathbf{V}_{\mathrm{out}}=\mathbf{T}\mathbf{V}_{\mathrm{in}}\mathbf{T}^\top+\mathbf{N},
\end{equation}
where $\mathbf{T}$ and $\mathbf{N}$ are  real (symmetric) $2\times2$ matrices. Eq.~(\ref{eq:CMevolution}) represents a channel if the following necessary and sufficient conditions are satisfied~\cite{schafer2013equivalence}:
\begin{equation}\label{eq:CPcondition}
  y\geq|\tau-1|,\hspace{5mm}\mathbf{N}\geq0,
\end{equation}
where $y\df\sqrt{\det{\mathbf N}}$ and $\tau\df\det{\mathbf T}$. In that case, the two remaining parameters characterizing all OMG channels are $\tau$ and $r=\min{[\rank{\mathbf T},\rank{\mathbf N}]}$.

\begin{figure}
  \resizebox{12.5cm}{!}{\includegraphics{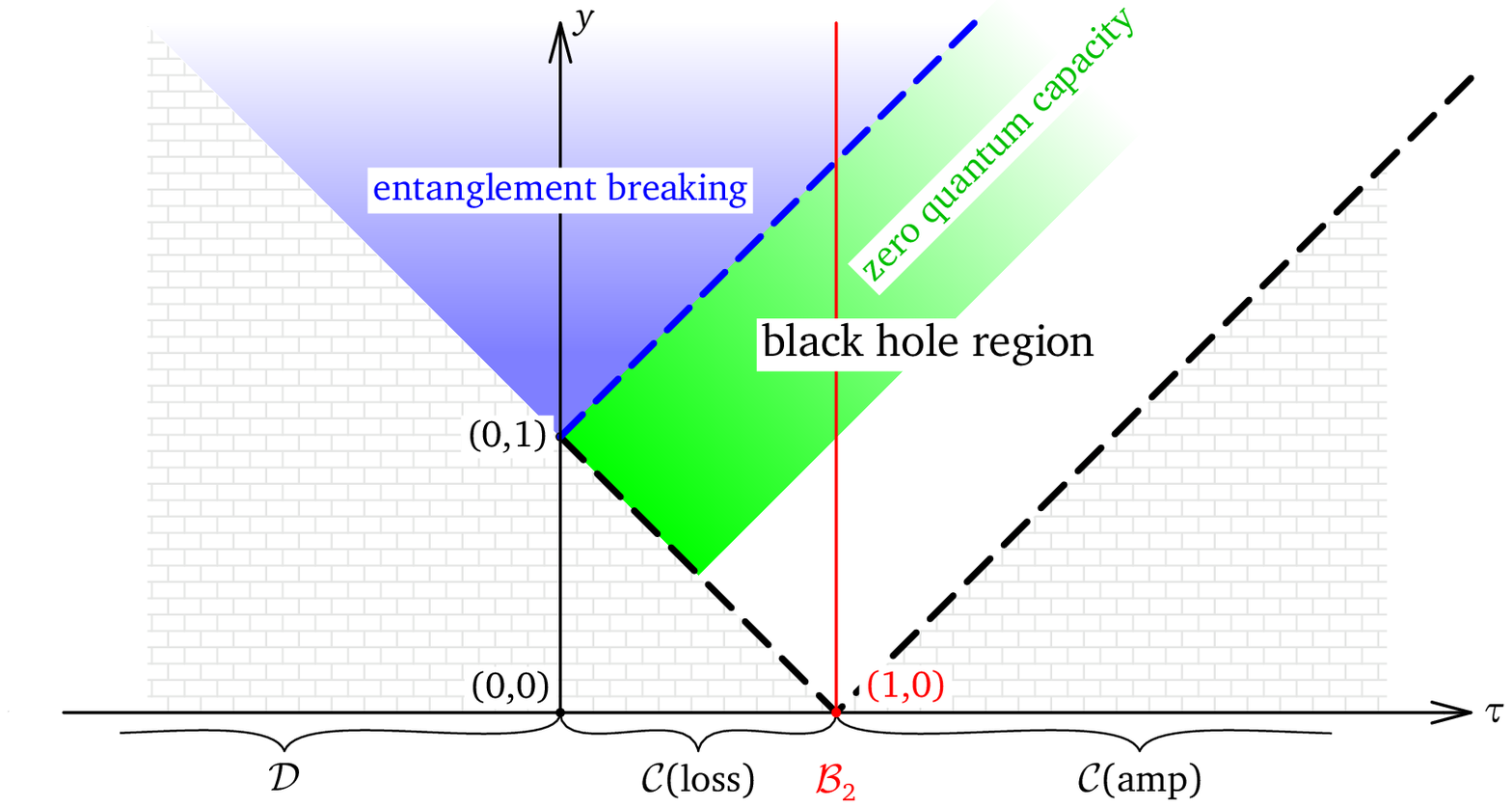}}
   \caption{The set of all rank-two OMG channels Eq.~(\ref{eq:CMevolution}) is parametrized by the pair $(\tau,y)$ given by $\tau=\det{\mathbf T}$ and $y=\sqrt{\det{\mathbf N}}$ and consists of lossy channels $\Ccal(\mathrm{loss})$, amplifiers $\Ccal(\mathrm{amp})$, conjugated channels $\Dcal$  and classical-noise channels ($\Bcal_2$ indicated by a red vertical line including the red dot: a rank-one class $\Bcal_2(\id)$ at $(1,0)$). The brickwall represents the area of non-complete positive maps where Eq.~(\ref{eq:CPcondition}) is violated (the unphysical region). The blue region contains entanglement-breaking channels for which $y\geq|\tau|+1$ holds~\cite{holevo2007one}, which are a subset of all zero quantum capacity channels (the green area covering the region $y\geq\tau$ intersects with all completely positive maps~\cite{caruso2006one}).
   For the black hole scenario we find that all Gaussian black hole channels (the $a$ mode) generating the outgoing radiation are confined to a semi-infinite strip demarcated by the dashed boundary (inclusive).}
    \label{fig:tauy}
\end{figure}
The parameters just defined can be used to write down the canonical representatives of all equivalence classes of OMG channels~\cite{holevo2007one,weedbrook2012gaussian}. If we were able to deduce a canonical form from the action of $S$ in Eq.~(\ref{eq:Smatrix}) we could find the form of a black hole Gaussian channel for all admissible $\a,\b,\g$ from Eq.~(\ref{eq:Bogoliubov}). This is indeed the case.

Consider an input three-mode state $\mathbf{V}_{abc,\mathrm{in}}=\id_a\oplus\id_b\oplus \mathbf{V}_{c,\mathrm{in}}$ where $\id$ stands for a two-dimensional identity matrix and
\begin{equation}\label{eq:Vin}
  \mathbf{V}_{c,\mathrm{in}}=
  \begin{bmatrix}
    e & g \\
    g & f \\
  \end{bmatrix}
\end{equation}
is a generic input OMG covariance matrix ($e,f,g\in\bbR\mbox{ s.t. }\mathbf{V}_{c,\mathrm{in}}+i\om\geq0$~\cite{simon1994quantum}). We then find that the Gaussian black hole channel is already in a canonical form where the matrices $\mathbf{N}$ and $\mathbf{T}$ are proportional to identity matrices resulting in:
\begin{equation}\label{eq:VaOut}
  \mathbf{V}_{a,\mathrm{out}}=
  \begin{bmatrix}
    e{s^2\over r^2}\sin^2{r}+\cos{2r}+{s^2\over r^2}\sin^2{r} & g{s^2\over r^2}\sin^2{r} \\
    g{s^2\over r^2}\sin^2{r} & f{s^2\over r^2}\sin^2{r}+\cos{2r}+{s^2\over r^2}\sin^2{r} \\
  \end{bmatrix}.
\end{equation}
This leaves us with only three possible candidates labeled in Ref.~\cite{weedbrook2012gaussian} as $\Ccal(\mathrm{loss})$,  $\Ccal(\mathrm{amp})$ and $\Bcal_2$  that possess the following canonical forms:
\begin{subequations}\label{eq:canForms}
  \begin{align}\label{eq:canForms1}
    (\mathbf{T},\mathbf{N})_{\Ccal(\mathrm{loss})} & = \big(\sqrt{\tau}\id,\ (1-\tau)(2\ave{n}+1)\id\big),\\
    (\mathbf{T},\mathbf{N})_{\Ccal(\mathrm{amp})} & = \big(\sqrt{\tau}\id,\ (\tau-1)(2\ave{n}+1)\id\big),\\
    (\mathbf{T},\mathbf{N})_{\Bcal_2} & = \big(\id,\ave{n}\id\big).
  \end{align}
\end{subequations}
The admissible values of $\tau$ are (in order):  $0\leq\tau<1,\tau>1$ and $\tau=1$.

The class $\Ccal(\mathrm{loss})$ represents lossy channels (for example the action of an unbalanced beam-splitter) and $\Ccal(\mathrm{amp})$ denotes the class of amplification channels. $\Bcal_2$ forms its own equivalence class of the so-called classical-noise channels~\cite{holevo2007one}. In the following we will refer to the family of channel Eqs.~(\ref{eq:canForms}) as the \emph{ Gaussian black hole channels}, defined by the $a$ subsystem of (\ref{eq:VaOut}). Using that equation we identify
\begin{subequations}\label{eq:tauANDy}
  \begin{align}
    \tau_a & ={s^2\over r^2}\sin^2{r}, \label{eq:tauANDy1}\\
    y_a & = \cos{2r}+{s^2\over r^2}\sin^2{r}\;,\label{eq:tauANDy2}
  \end{align}
\end{subequations}
where we have introduced the subscript $a$ to the parameters $\tau$ and $y$ to indicate that the $a$ subsystem is the quantum black hole channel output. We further verify that the first condition of complete positivity in Eq.~(\ref{eq:CPcondition}) is satisfied whenever $s^2\geq r^2$. The second condition $N\geq0$ gives a weaker constraint.

How large is the set of channels given by~(\ref{eq:tauANDy}) with respect to the classes it is part of? To visualize the set we adopt a figure from~\cite{schafer2013equivalence,giovannetti2013ultimate} where all (four) rank $r=2$ OMG channels are parametrized using the coordinates $(\tau,y)$, see Fig.~\ref{fig:tauy}. The form of Eq.~(\ref{eq:tauANDy1}) dictates $\tau_a\geq0$. Consequently, from~(\ref{eq:tauANDy2}) we obtain $y_a=\cos{2r}+\tau_a$ and this leads to an interesting observation: If we choose a given $\tau_a\geq0$, then by adjusting $r$ (and therefore $s\geq0$ to keep $\tau_a$ constant) the value of $y_a$ oscillates between $\tau_a-1$ and $\tau_a+1$ for all $\tau_a$. The oscillation boundaries become  $|\tau_a-1|$ and $\tau_a+1$ for $0\leq\tau_a<1$ due to the condition of complete positivity.

Among all OMG channels, many are entanglement breaking (blue area in Fig.~\ref{fig:tauy}). Recall that a quantum channel $\Ncal$ is entanglement breaking if  the state $(\id_A\otimes\,\Ncal_B)(\vr_{AB})$ is separable (i.e. classically correlated) for any bipartite entangled state $\vr_{AB}$. The black hole channels that lie in the area marked ``black hole region'', on the other hand, are \emph{ not} entanglement breaking: they are composed of the equivalence classes $\Ccal(\mathrm{loss})$, $\Ccal(\mathrm{amp})$ and $\Bcal_2$. The only exception to this rule is part of the boundary region $y=\tau+1$ ($\tau\geq0$) denoted by the dashed blue line, where even the black hole channel is entanglement breaking.

It is instructive to investigate what channel in the black hole region in Fig.~\ref{fig:tauy} we obtain if we choose certain limiting parameters of the Bogoliubov transformation~Eq.~(\ref{eq:Bogoliubov}). For example, as the parameter $\a$ sets the reflectivity of the black hole, we could study $\a=0$ and $\a=1$ as they correspond to the limiting cases of a perfectly reflecting and absorbing black hole, respectively.~\cite{AdamiVersteeg2013,bradler2013capacity}. Indeed, these two cases were studied in~\cite{bradler2013capacity} for the case of a qubit black hole channel. To obtain $\a=1$ we set $r=0$ and from Eqs.~(\ref{eq:tauANDy}) we obtain $\tau_a=s^2$ and $y_a=1+s^2$, corresponding to the blue dashed line in Fig.~\ref{fig:tauy} (recall that $s\geq r=0$) and this is precisely the \emph{ only} instance where the black hole channel is entanglement breaking (for all $s$). This is consistent with the (non-Gaussian) dual-rail encoding studied in~\cite{bradler2013capacity} where we found the case $\a=1$ to be entanglement breaking as well. For the other case $\a=0$ we set $r=\pi/2$ and this time we find $\tau_a=4s^2/\pi^2$ and $y_a=4s^2/\pi^2-1$. Because $s\geq\pi/2$ we obtain $\tau_a\geq1$ and $y_a\geq0$ and we can identify the channel region to be the lower semi-infinite dashed boundary in Fig.~\ref{fig:tauy}, corresponding to degradable channels (see Sec.~\ref{subsec:directcode}). But this finding is again in perfect agreement with the (physically very different) qubit case we studied earlier~\cite{bradler2013capacity}, where we also found that the perfectly reflecting channel (known also as the Unruh channel~\cite{bradler2011infinite}) is degradable. As a matter of fact, all values $r\in[0,\pi/2]$ generate  half-lines that ``foliate'' the semi-infinite black hole strip in~Fig.~\ref{fig:tauy}.

\section{Quantum Capacity of OMG Channels}

The classical capacity of phase-insensitive OMG channels was studied by Giovannetti et al.~\cite{giovannetti2013ultimate}, who derived explicit expressions for them. Here, we are interested in calculating the \emph{ quantum} capacity of these channels, and in particular the quantum capacity of the black hole channels depicted in Fig.~\ref{fig:tauy}. Unlike their classical capacities, the quantum capacity can only be explicitly calculated for a small fraction of channels because a single-letter formula is not presently known for all~\cite{holevo2012quantum}.

A large area of parameter space of the black hole channels (the green area in Fig.~\ref{fig:tauy}) has vanishing quantum capacity~\cite{caruso2006one}, owing to the fact that these channels can be written as a composition of an arbitrary Gaussian channel and an anti-degradable Gaussian channel~\cite{holevo2012quantum}.
The partial overlap of the Gaussian black hole channels with the zero quantum capacity region is fundamental from a physical point of view, because this overlap implies that for some black holes (that is, some values of $r$ and $s$ in the evolution operator Eq.~(\ref{eq:Smatrix}), quantum information cannot be recovered by an outside observer.

\begin{figure}[t]
\centering
  \includegraphics{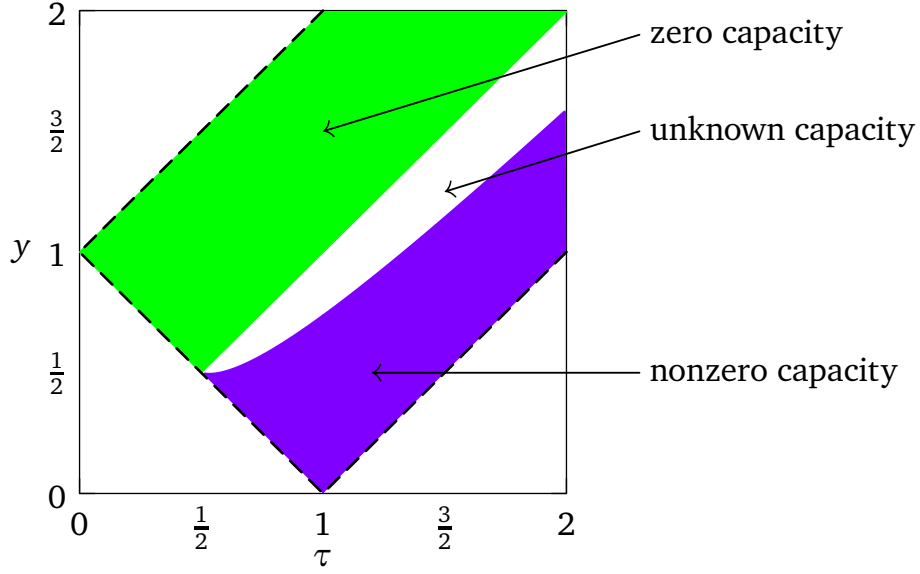}
   \caption{The black hole region (the green and white area within the dashed lines in Fig.~\ref{fig:tauy}), colored to emphasize the (calculable) non-zero values of the coherent information (purple area) from Eq.~(\ref{eq:cohinfo}). This expression gives a non-zero lower bound for the quantum capacity of the Gaussian black hole channel. The green area represents a region with zero quantum capacity, as discussed earlier. The white region in between is uncharted territory, that is, where the quantum capacity is unknown. The coherent information is zero in this region, but this does not imply that the quantum capacity vanishes, as the coherent information is just a lower bound~\cite{holevo2001evaluating}.}
    \label{fig:cohinfo}
\end{figure}

The situation is more complicated in the white part of the black hole region of Fig.~\ref{fig:tauy}, where the capacity cannot so easily be calculated. In Fig.~\ref{fig:cohinfo}, we focus only on the black hole region of the OMG channel parameter space.  We remind the reader that the black dashed line that separates the green area in Fig.~\ref{fig:cohinfo} from the entanglement breaking channels (the blue area in Fig.~\ref{fig:tauy}) itself describes entanglement breaking channels, which in fact are known to have vanishing capacity~\cite{holevo2012quantum}. The black dashed line from $(0.5,0.5)$ to $(0,1)$ separating the green area from the unphysical maps consists of channels that are anti-degradable, and therefore their capacity also vanishes. The remaining area of black hole channels (the white area in Fig.~\ref{fig:tauy}, also the white and purple area in Fig.~\ref{fig:cohinfo}) contains channels with both calculable and unknown capacities.

On a part of the boundary of that region (depicted by the black dashed lines from $(1,0)$ to $(0.5,0.5)$ and $y=\tau-1$) the OMG channels are known to be degradable~\cite{wolf2007quantum,caruso2006degradability}, which implies that the capacity is calculable (and thus known). Inside of this area the quantum capacity can only be bounded from below~\cite{holevo2001evaluating,holevo2012quantum}. Moreover, it is not known from general principles whether the quantum-capacity achieving codes in this area are Gaussian at all, unlike in the boundary (degradable) region where the optimal quantum codes are coherent (and therefore Gaussian) states~\cite{wolf2007quantum}.

Our starting point for calculating the quantum capacity is the optimized coherent information of a general quantum channel $\Ncal$ introduced in Eq.~(\ref{eq:singlecopyQcap}). Using~(\ref{eq:DEFcohinfo}) and the definition of the complementary channel $\wh\Ncal$ we can write it as
\begin{equation}\label{eq:cohInfoGeneral}
I_{\mathrm{coh}}(\Ncal)=\sup_{\vr}{[H(\Ncal(\vr))-H(\wh{\Ncal}(\vr))]}.
\end{equation}
For a special case of OMG channels (labeled by $\Gcal$)  Eq.~(\ref{eq:cohInfoGeneral}) becomes~\cite{holevo2001evaluating,holevo2012quantum}
\begin{equation}\label{eq:Icoh}
  I_{\mathrm{coh}}(\Gcal)\df\sup_{N}{\tilde{I}_{\mathrm{coh}}(N,\Gcal(\vr))}=\sup_{N}{\big[g(N')-g(x_+)-g(x_-)\big]},
\end{equation}
where $N=\Tr{}[\vr\,a^\dg a]$ is the mean particle number of an input Gaussian state $\vr$ and $g(x)$ is its von Neumann entropy~\cite{agarwal}
\begin{equation}\label{eq:g}
g(x)\equiv(1+x)\log{(1+x)}-x\log{x}\;.
\end{equation}
Following Refs.~\cite{holevo2001evaluating,holevo2012quantum}, we set
\begin{subequations}\label{eq:cohInfoPars}
\begin{align}
  N' & = \begin{cases}\label{eq:cohInfoParsA}
        \tau N+K         \quad & \mbox{for}\ 0\leq\tau<1\\
        \tau N+\tau-1+K  \quad & \mbox{for}\ \tau>1\,,
  \end{cases}\\
  x_+ & = {1\over2}(D+N'-N-1),\label{eq:cohInfoParsB}\\
  x_- & = {1\over2}(D-N'+N-1),\label{eq:cohInfoParsC}
\end{align}
\end{subequations}
where
\begin{align}
  K&={1\over2}(y-|1-\tau|)\label{eq:K}, \\
  D&=\big((N+N'+1)^2-\tau\,4N(N+1)\big)^{1/2}.\label{eq:D}
\end{align}
The first line of Eq.~(\ref{eq:cohInfoPars}) has a neat interpretation of the added classical noise $K$ present in the environment plus the mean photon number in the channel input $N$ modulated by the transmissivity factor $\tau$. The second line has a similar meaning except that now the parameter $\tau>1$ is related to the amplifier gain and the component $\tau-1$ is a vacuum contribution. The expressions in Eqs.~(\ref{eq:cohInfoParsB}) and~(\ref{eq:cohInfoParsC}) are eigenvalues of certain matrices whose entropy is needed to be calculated in order to find the coherent information.

In general, the optimized coherent information is maximized over all possible codewords, here restricted to Gaussian states~\cite{holevo2012quantum}. Unlike in the calculation of the classical capacity~\cite{holevo2001evaluating}, the limit of infinite input power $N\to \infty$ does not \emph{ usually} lead to a diverging entropic quantity. It turns out that for the value $K=0$, the maximization over codewords is achieved for $N\to\infty$ for $\tau>1/2$ because  Eq.~(\ref{eq:Icoh}) is increasing monotonically. For $\tau<1/2$ the function decreases monotonically, and the maximal value is reached for $N=0$, leading to a vanishing coherent information. This (vanishing) value is compatible with the zero quantum information in the green region in Fig.~\ref{fig:cohinfo})~\cite{holevo2012quantum}.

When $K>0$ the coherent information does not increase monotonically for $\tau>1/2$ as a function of $N$. One of the authors recently showed that even in this case the supremum is achieved for $N\to\infty$~\cite{bradler2015coherent}. Taking the limit, we can generalize the coherent information expression and obtain:
  \begin{align}\label{eq:cohinfo}
   I_{\mathrm{coh}}(\Gcal(\vr))
   &=\lim_{N\to\infty}\tilde{I}_{\mathrm{coh}}(N,\Gcal(\vr))\nn\\
   &={K\over|1-\tau|}\log{K\over|1-\tau|}-{|1-\tau|+K\over|1-\tau|}\log{|1-\tau|+K\over|1-\tau|}
   +\log{\tau\over|1-\tau|},
  \end{align}
where $\Gcal(\vr)$ is $\Ccal(\mathrm{loss})$ or $\Ccal(\mathrm{amp})$. The equation reduces to $I_{\mathrm{coh}}=\log{\tau\over|\tau-1|}$ for $K=0$, which coincides with the expression from~\cite{holevo2012quantum} (see also~\cite{wolf2007quantum}) valid whenever $\tau>1/2$. Indeed, $K$ vanishes on the dashed boundary of the purple region and of the green region (where $y<1$) in Fig.~\ref{fig:cohinfo}.  The dashed boundary of the purple region is where the quantum capacity is nonzero and actually calculable -- the OMG channels there are known to be degradable.

The case $\tau=1$ must be treated separately, as channels with $\tau=1$ represent a separate class of zero added classical noise channels (denoted by $\Bcal_2$, see Fig.~\ref{fig:tauy}). We find
$$
\lim_{\tau\to1^-}\lim_{N\to\infty}\tilde{I}_{\mathrm{coh}}(N,\Ccal(\mathrm{loss}))\equiv\lim_{\tau\to1^+}\lim_{N\to\infty}\tilde{I}_{\mathrm{coh}}(N,\Ccal(\mathrm{amp}))=-1-\log{K}\;.
$$
Evidently, this is a singular case where the optimized coherent information (and by inference the quantum capacity, see Eq.~(\ref{eq:QuantCaplowerbound})) diverges. This happens when $K=0$ (while $\tau=1$): the rather exceptional point $(1,0)$ in Fig.~\ref{fig:tauy}, corresponding to the subclass of $\Bcal_2$ channels where $\mathbf{T}=\id$ and $\mathbf{N}=0$ in Eq.~(\ref{eq:CMevolution})~\cite{holevo2007one}. In other words, this is a trivial noiseless (identity) Gaussian channel, whose capacity is known to diverge in classical physics~\cite{cover2012elements}.

\section{Complementary OMG channel}
\label{sec:complementarychannels}

We just observed that a sizable swath of the black hole OMG channel space has a vanishing capacity, implying that quantum information cannot be retrieved from it by an outside stationary observer. As discussed previously, this implies that the observer cannot perfectly reconstruct the quantum entanglement that the sender has been part of. However, as opposed to a loss of classical information that would imply the loss of microscopic time-reversal invariance~\cite{witten2012quantum}, the loss of quantum information does not contradict any known laws of physics. But we can nevertheless ask \emph{ where} the quantum information is hiding, because in a completely unitary picture of quantum dynamics,  quantum information cannot be lost from the universe. In this section we show that the quantum information is available to observers behind the horizon, by calculating a lower bound the capacity of the quantum channel to send information beyond the horizon: the complementary OMG channel.

Before we proceed, it is important to emphasize a particularity of the isometry $e^X$ from Eq.~(\ref{eq:ansatz}). So far, we studied the Gaussian black hole channels that correspond to the $a$ mode (the horizon mode available to outside observers). The minimal Stinespring dilation for these channels would usually be a two-mode isometry satisfying certain properties discussed in Ref.~\cite{weedbrook2012gaussian}. However, the expression $e^X$ is a \emph{ three-mode} operator, and this could conceivably not be the most economic dilation whose reference mode is in a pure state~\cite{caruso2011optimal}. For example,  on the black dashed boundary in Fig.~\ref{fig:tauy} the minimal Stinespring dilation is known to be a two-mode isometry~\cite{giovannetti2013ultimate}. On the other hand, our redundant (non-economic) isometry contains valuable information about the physics of the black hole interaction. Namely, the two modes forming the complementary black hole channel ($b$ and $c$) correspond to the physical system of the black hole itself and the radiation that crosses the event horizon, respectively, which are both integral to the channel.


Could it be that the green area for the $a$ mode (the black hole channel in Fig.~\ref{fig:tauy} or~\ref{fig:cohinfo} with vanishing capacity) corresponds to a \emph{non-zero} quantum capacity of the $c$ mode (the radiation penetrating the horizon from outside) and vice-versa? To study this, we must investigate the fate of the $c$ mode explicitly. Using the same input state $V_{abc,\mathrm{in}}$ as before, we find for the parameters that characterize the $c$-channel \emph{into} the black hole
\begin{subequations}\label{eq:VcOutParams}
\begin{align}
   \tau_{c}& =  \left(\frac{s^2 (\cos{r}-1)}{r^2}+1\right)^2, \\
   y_c & = {s^2\over r^2}\bigg({s^2\over r^2}-1\bigg)(\cos{r}-1)^2+{s^2\over r^2}\sin^2{r}.
\end{align}
\end{subequations}

\begin{figure}[t]
  \resizebox{12cm}{!}{\includegraphics{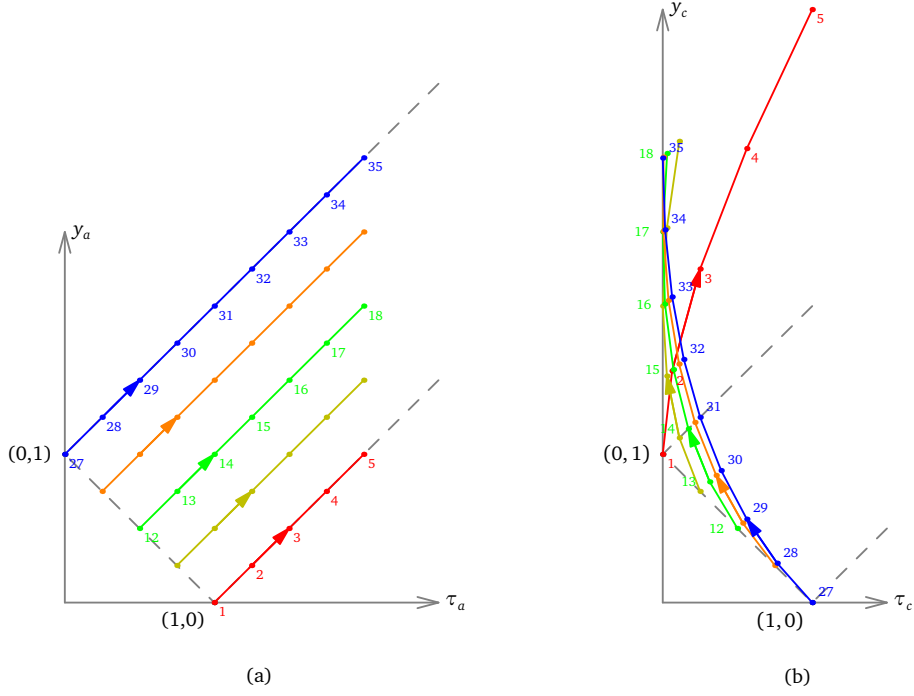}}
   \caption{OMG channel properties outside and inside the horizon. (a): We labelled 35 points in the space of parameters $(\tau_a,y_a)$ that characterize 35 OMG channels sampling the black hole region from Fig.~\ref{fig:tauy} -- the Gaussian black hole channel (mode $a$). The numbers in the parenthesis on the axes are the coordinates $(\tau_a,y_a)$. (b): Each of the $a$-channel parameters is transformed into the corresponding $c$-channel parameters as described in the text. By following the points (and lines) we can study the relationship between the $a$-mode and $c$-mode channels. It is important to keep in mind that (as discussed in the text) the $c$ mode is not the entire complementary subsystem to the black hole channel, see the discussion before Eq.~(\ref{eq:conjchannel}). The numbers in the parenthesis on the axes are the coordinates $(\tau_c,y_c)$.
      }
    \label{fig:for}
\end{figure}
We can express $\tau_c$ and $y_c$  in terms of $\tau_a$ and $y_a$ by first writing (using Eqs.~(\ref{eq:tauANDy}))
\begin{equation}\label{eq:rFunction}
  r={1\over2}\acos{(y_a-\tau_a)}+k\pi,
\end{equation}
where $k\in\bbZ$. It then from Eqs.~(\ref{eq:tauANDy}) and (\ref{eq:VcOutParams}) that
\begin{subequations}\label{eq:VcOutParamsIntermsOfa}
\begin{align}\label{eq:VcOutParamsIntermsOfa1}
   \tau_{c}& =  \left(\frac{\tau_a}{\sin^2{r}}(\cos{r}-1)+1\right)^2, \\
   y_c & = \frac{\tau_a}{\sin^2{r}}\bigg(\frac{\tau_a}{\sin^2{r}}-1\bigg)(\cos{r}-1)^2+\tau_a.
\end{align}
\end{subequations}
We can now see what the corresponding OMG channel in the $c$ mode is, for a given $a$ mode channel. It turns out that for a given pair $(\tau_a,y_a)$ inside the black hole parameter region (that is, excluding the dashed boundary in Fig.~\ref{fig:tauy}) we find two solutions $(\tau_c,y_c)$ corresponding to even or odd $k$ in Eq.~(\ref{eq:rFunction}). In other words, the map from the set of all admissible parameters to the set of all black hole channels is onto but not one-to-one (it is actually two-to-one quotiented by $2\pi$). This two-to-one mapping deserves a closer look.

The physical properties of a black hole are determined by the isometry parameters $r$ and $s$ in Eqs.~(\ref{eq:abg}). They uniquely specify the isometry $e^X$ and therefore the Gaussian black hole channel (the $a$ output). Suppose that another pair of parameters $r',s'$ that leads to a different isometry $e^{X'}$ induces the same Gaussian black hole channel (this is in principle possible). This implies that the complementary channel (the $bc$ output) induced by the those two isometries must also be equivalent. However, it does \emph{ not} imply this identity for any \emph{ part} of the complementary channel (the $b$ or $c$ outputs alone). That is precisely what is happening here: a black hole channel defined by $(\tau_a,y_a)$ leads to \emph{ two} solutions of the black hole parameters $r,s$ in Eq.~(\ref{eq:rFunction}) for the $c$-channel.

The first solution  (even $k$) is depicted in Fig.~\ref{fig:for}(a) where the semi-infinite rectangle from Fig.~\ref{fig:tauy} is sampled by 35 points in the black hole region close to the origin. The corresponding $c$-mode channel parameters are shown in Fig.~\ref{fig:for}(b).
We can see that there are instances where both channels have vanishing capacity (for example, the set of parameters 30). The only region where there is a certain kind of complementarity between the $a$ and $c$ modes is the line connecting the points $(0,1)$ and $(1,0)$ (note the opposite orientation of this line in both panels), and also the red line covering degradable channels (if continued indefinitely).

\begin{figure}[t]
  \resizebox{13cm}{!}{\includegraphics{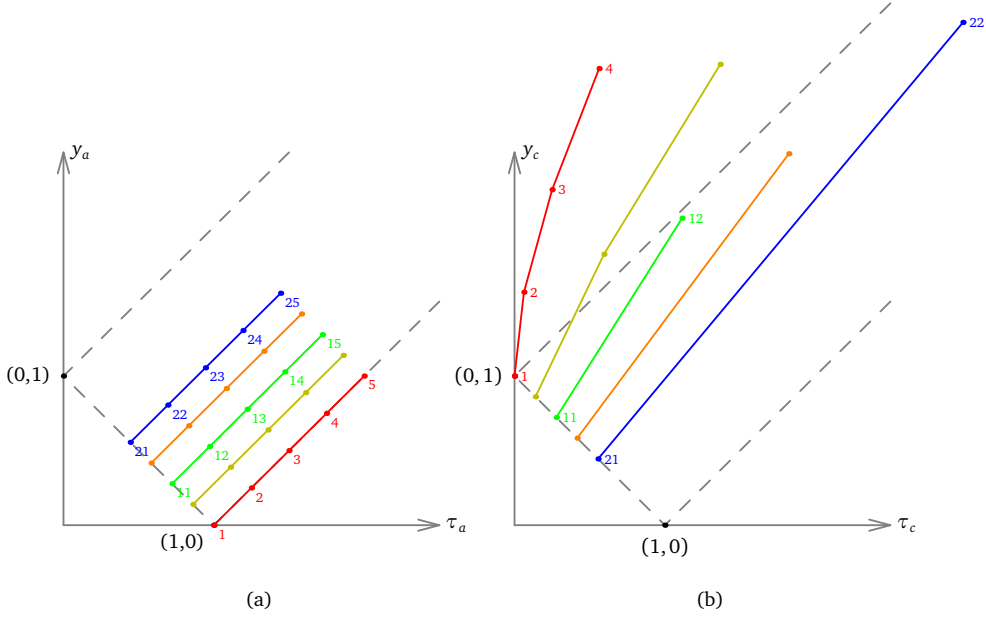}}
   \caption{Correspondence between $a$-mode and $c$-mode channels for odd $k$. $\tau_c$ and $y_c$ in Eq.~(\ref{eq:VcOutParamsIntermsOfa}). (a): Sample of 25 channels in the black hole region and (b): their corresponding $c$-mode channels. The missing channel numbers in panel (b) are too far from the origin to appear in the figure.}
    \label{fig:for2}
\end{figure}

The solutions corresponding to odd $k$ are depicted in Fig.~\ref{fig:for2}(b). We sampled the black hole region Fig.~\ref{fig:for}(a) slightly differently with 25 points compared to $k$ even.
We stress again that the existence of two different $c$ channels corresponding to a given Gaussian black hole channel is not surprising because the $c$ channel is not the whole complementary channel to the black hole channel.

By analyzing $\mathbf{V}_{b,\mathrm{out}}$ from Eq.~(\ref{eq:CMevolution}) we find that for all Gaussian input states, the $b$ output subsystem is given by an OMG channel from the conjugated equivalence class $\Dcal$, whose generic form reads~\cite{holevo2007one,weedbrook2012gaussian}
\begin{equation}\label{eq:conjchannel}
  (T,N)_{\Dcal}  = \big(\sqrt{-\tau}\s_z,\ (1-\tau)(2\ave{n}+1)\id\big),
\end{equation}
where $\tau<0$ and $\s_z$ is the Pauli $z$ matrix. Inspecting Fig.~\ref{fig:tauy}, we see that all these channels lie in the entanglement breaking region and by themselves cannot carry any quantum information. But if the information is never in the $b$ mode and often neither in the $c$ or $a$ modes, does it mean it is lost? It turns out that this is not necessarily the case.

As we stressed several times, the channel across the black hole horizon (the $c$-mode channel) is not the full black hole complementary channel, which must be ``spanned'' by both the $b$ and $c$ modes. Because this is a two-mode channel, we do not know to what Gaussian channel it corresponds to (the present analysis is concerned exclusively with OMG channels). However, this does not prevent us from estimating its quantum capacity.
The green region in Figs.~\ref{fig:tauy} and~\ref{fig:cohinfo} is a zero capacity region and so $I_{\mathrm{coh}}(\Gcal)=Q(\Gcal)=0$, where $\Gcal$ is one of the identified Gaussian channels~in Eq.~(\ref{eq:canForms}). The quantum capacity is zero due to the fact that the green channels are antidegradable~\cite{caruso2006one}. Therefore the only Gaussian codewords $\vr$ that exist are such that $H(\Gcal(\vr))-H(\wh{\Gcal}(\vr))$ is negative and the supremum is necessarily  zero.

The very same Gaussian codes are then used to calculate the coherent information for the Gaussian complementary channel $\wh\Gcal(\vr)$ (the $bc$ modes) and naturally they give a positive value even without taking the supremum
\begin{equation}
 I_{\mathrm{coh}}(\wh{\Gcal})>H(\wh\Gcal(\vr))-H({\Gcal}(\vr))>0.
\end{equation}
As a consequence, the supremum itself is positive for the complementary channel. We may conclude that the quantum capacity $Q(\wh\Gcal)\geq I_{\mathrm{coh}}(\wh{\Gcal})$ (see Eq.~(\ref{eq:QuantCaplowerbound})) is positive in general (even if it vanishes individually for the $c$ and $b$ modes), and the information is delocalized in the modes inside the black hole whenever the Gaussian black hole channel capacity outside vanishes (the green region in Fig.~\ref{fig:tauy} or \ref{fig:cohinfo}). As a matter of fact, we can say more. Since the zero capacity green region is formed by channels $\Gcal$ that are Gaussian antidegradable, the complementary channels $\wh\Gcal$ are degradable as discussed in Sec~\ref{subsec:directcode}. Therefore the capacity is given by the single-letter expression Eq.~(\ref{eq:singleletter}): $Q(\wh\Gcal)=I_{\mathrm{coh}}(\wh{\Gcal})$~\cite{holevo2012quantum}. Finally, notice that the only \emph{known} point in the dashed boundary region in Fig.~\ref{fig:cohinfo}, where the channel is both degradable and antidegradable, is the one with the coordinates $(\tau,y)=(1/2,1/2)$. There is a big question mark of what happens in the white region.

\section{Conclusions}

We analyzed the late time interaction of a scalar field in the form of Gaussian states with a Schwarzschild black hole based on Sorkin's model. For a distant outside observer such a black hole acts as a quantum channel from a family of one-mode Gaussian channels that were recently classified.  Here, we refer to these channels as  ``Gaussian black hole'' channels. The classification enables us to ask the following question:  how much information can a distant observer at future infinity recover if a Gaussian state carrying classical or quantum information interacts with an already formed black hole?

This question is nothing but a reformulation of the black hole information loss problem, and it can be answered by a calculation of the classical and quantum capacity of these quantum Gaussian channels. The classical capacity of phase-insensitive one-mode Gaussian channels was calculated by Giovannetti et al.~\cite{giovannetti2013ultimate} but it is the fate of quantum information that is arguably more relevant for the black hole information puzzle. By mapping the black hole to the Gaussian quantum channels in this manner, we are closer to resolving the fate of information interacting with a black hole. In particular, we  found the exact parameter region of one-mode Gaussian channels corresponding to an arbitrary Schwarzschild black hole. These black hole Gaussian channels represent an interesting subset of three equivalence classes of Gaussian channels that are (with one exception) non-entanglement breaking.  We find that half of the parameter space of black hole channels has vanishing quantum capacity. The other half of this region splits into two parts: one part where the quantum capacity is calculable (or is at least known to be nonzero) and the other where no non-negative lower bound is known. There is, however, currently no argument that excludes a positive quantum capacity for those channels.

We also studied the complementary channel (sending quantum information on the other side of the black hole horizon) and found that the channel's capacity is given by the optimized coherent information, and is positive in one half of the black hole parameter region when the capacity to transmit quantum information outside the black hole vanishes.
In this manner, the no-cloning theorem is respected (as also seen in the dual-rail channel discussed in~\cite{bradler2013capacity}): quantum information is never available in two places. At the same time, each of the modes $b$ and $c$ that together compose the complementary channel may not be sufficient to reconstruct quantum states inside the black hole by themselves.

What are the consequences for (quantum) information retrieval from a black hole? That is a question we can answer only partially. We are not aware of any physical mechanisms restricting or favoring some sections of the whole black hole channel parameter region, which implies that quantum information can be lost within black holes. We can only say that if the black hole resides in the nonzero capacity region, quantum information can be retrieved with arbitrary accuracy, and we can calculate or estimate the rate of recovery (certain extreme points such as the noiseless identity channel $\Bcal_2(\id)$ can be safely ignored, as the black hole always introduces some noise due to the presence of spontaneous emission of radiation, namely the Hawking radiation effect). If on the other hand a black hole channel is in the zero capacity region, quantum information cannot be recovered from it on the outside. This, however, does not imply a breakdown of quantum mechanics, or any other of our known laws of physics.
Thus, it appears that mapping black holes to a one-mode Gaussian channel allows us to understand how black holes process classical or quantum information using concepts from quantum optics and quantum information theory only.

\section*{Acknowledgement}
\thanks{The authors thank Oleg Pilyavets for comments.}

\bibliographystyle{unsrt}


\end{document}